\documentclass[twocolumn]{jpsj3}
\usepackage{newtxtext}
\usepackage[varg]{newtxmath}
\usepackage{braket}
\usepackage{bm}
\usepackage[normalem]{ulem}

\usepackage{graphicx}
\usepackage{dcolumn}
\usepackage{bm}
\usepackage{amsmath}
\usepackage[usenames]{color}

\DeclareMathOperator{\sgn}{sgn}

\newcommand{\journal}[5]{{#1}, {#2}\ \textbf{#3}, {#4} ({#5}).}
\newcommand{\arxiv}[2]{{#1}, arXiv:{#2}.}
\def\PR{Phys.\ Rev.}
\def\PRB{Phys.\ Rev.\ B}
\def\PRL{Phys.\ Rev.\ Lett.}

\def\JPSJ{J.\ Phys.\ Soc.\ Jpn.}
\def\RMP{Rev.\ Mod.\ Phys.}
\def\EPL{Europhys.\ Lett.}

\def\SciRep{Sci.\ Rep.}
\def\OptExp{Opt.\ Express}
\def\JCP{J.\ Chem.\ Phys.}
\def\RPP{Rep.\ Prog.\ Phys.}
\def\AP{Adv.\ Phys.}

\title{Photoinduced topological spin texture in a metallic ferromagnet}
\author{Atsushi Ono\thanks{ono@cmpt.phys.tohoku.ac.jp} and Sumio Ishihara}
\inst{Department of Physics, Tohoku University, Sendai 980-8578, Japan} 

\abst{
Photoinduced nonequilibrium spin structure is examined in the double-exchange model, in which itinerant electrons couple with localized spins through the ferromagnetic Hund coupling.
In particular, we focus on the transient spin structure from the initial ferromagnetic metallic state to the steady antiferromagnetic ordered state reported in [Phys.\ Rev.\ Lett.\ \textbf{119}, 207202 (2017)].
By solving the Schr\"odinger equation combined with the Landau-Lifshitz-Gilbert equation,
we find finite winding number and chirality, which implies emergence of topological chiral spin textures.
These observations are reproduced by a calculation where spin dynamics after sudden quench of the chemical potential are examined in larger clusters.
A possible mechanism of the topological spin texture in the transient dynamics is discussed.
}

\begin{document}
\maketitle


Topological character in magnetic materials is one of the attracting scientific themes in recent condensed matter physics.~\cite{zang,hellman,togawa,batista}
Chiral-spin soliton, spin vortex and skyrmion are the examples of the topological spin textures in one-, two- and three-dimensional magnetic solids, respectively.
These objects usually emerge in thermally and magnetically excited states in ferromagnetic (FM) or antiferromagnetic (AFM) ordered states, and persist due to the topological protection.
In addition, the chiral spin orders, e.g., spin spiral, conical and screw structures, are widely recognized as non-trivial spin states, which often contribute to the functions such as multiferroics and the anomalous Hall effect.
One of the strategies to produce the topological spin objects is to utilize the relativistic spin-orbit interaction (SOI).
This provides the spin anisotropy in crystalline lattice, and often induces non-collinear or non-coplaner spin structures through the antisymmetric exchange interaction.
However, this strategy is limited to the materials including magnetic elements with large SOI.

Intense laser-light irradiation to solids is now widely known to induce new states of matter in strong nonequilibrium states, in which novel electronic and lattice structures unrealized in thermal equilibrium states are expected.~\cite{kirilyuk,tokura,aoki}
Pulse-laser-induced superconductivity~\cite{fausti,mitrano} is one of the examples of the photoinduced novel states.
Laser light is also available to control the topological spin textures, for examples,
the electromagnon through the inverse Dzyaloshinskii-Moriya interaction,~\cite{mochizuki} the skyrmion by using the circularly polarized light~\cite{ogawa,stepanov} or the optical vortex~\cite{fujita,yang} and so on.

In this Letter, we study the topological spin textures in the photoinduced nonequilibrium state in a metallic magnetic system.
We adopt the double-exchange (DE) model,~\cite{zener,anderson,degennes} which has been applied to correlated electron systems, metallic magnets and spintronics related phenomena.
Theoretical studies of the DE model revealed the photoinduced transition from the AFM insulating to FM metallic states.~\cite{chovan,matsueda,kanamori,ohara,koshibae1,koshibae2}
In the previous paper by the authors~\cite{ono1,ono2}, it was found that the FM metallic state described by the DE model is transformed into the AFM state by photoirradiation.
Here, we focus on the transient spin structure described by the DE model from the initial FM state to the AFM ordered state.
We find emergence of the topological spin texture in the photoinduced transient state through the calculation of the real-space spin configuration, the Pontryagin index, and the scalar chirality.
This observation is reproduced by analyzing the spin dynamics after chemical-potential quench in a large cluster.
Relations to the topological defect production in long-range ordered states are discussed.

We adopt the DE model defined by
\begin{align}
{\cal H} = -h \sum_{\langle ij \rangle s} c_{is}^\dagger c_{js} - \frac{J}{S} \sum_{iss'} \bm{S}_i \cdot \bm{\sigma}_{ss'} c_{is}^\dagger c_{is'},
\label{eq:hamiltonian}
\end{align}
where $c_{is}^\dagger \ (c_{is})$ is a creation (annihilation) operator of a conduction electron with spin $s \ (={\uparrow},{\downarrow})$ at site $i$, $\bm{S}_i$ is a localized-spin operator with magnitude $S$, and $\sigma^\alpha \ (\alpha=x,y,z)$ are the Pauli matrices.
The first term describes the electron hopping, and the second term represents the on-site Hund coupling with $J>0$.
We adopt the finite-size clusters with the total number of sites $N$, that of electrons $N_e$, and the electron number density $n_e=N_e/N$.
From now on, we consider the case in which $J/S=4h$ and $n_e=0.5$, which provides the FM metallic ground state.~\cite{yunoki}

The electromagnetic field is introduced in the Hamiltonian as the Peierls phase as $h c_{is}^\dagger c_{js} \rightarrow h \exp[i \bm{A}(t)(\bm{r}_i-\bm{r}_j)] c_{is}^\dagger c_{js}$, where $\bm{A}(t)$ is the vector potential at time $t$ and $\bm{r}_i$ is a position of the $i$th site.
We consider the continuous-wave (cw) field given by $\bm{A}({t}) = (\bm{F}_0/\mathit{\Omega}) \sin(\mathit{\Omega}{t})$, where $\bm{F}_0$ and $\mathit{\Omega}$ are amplitude and frequency of the electric field, respectively.
We adopt the two-dimensional square lattice with the lattice constant $a$.
The electric field is applied to the diagonal direction, $\bm{F}_0=(F_0,F_0)$.
Energy and time are measured in units of $h$ and $\hbar/h$, respectively.
The nearest-neighbor hopping amplitude $h$, the reduced Planck constant $\hbar$, the electron charge, and the lattice constant are taken to be unity.

The real-time dynamics of electrons and spins are calculated by the method adopted in Refs.~\citen{koshibae1,koshibae2,ono1}.
Here, we summarize briefly the calculation procedure.
The localized spins are treated as classical vectors, which is justified in the limit of large $S$.
The Hamiltonian at time $t$ is diagonalized as ${\cal H}({t})=\sum_{\nu} \varepsilon_\nu({t}) \phi_\nu^\dagger({t}) \phi_\nu({t})$, where $\phi_\nu^\dagger(t)$ is a creation operator of an electron with energy $\varepsilon_\nu$.
The wavefunction at time ${t}$ is given as a single Slater determinant $\vert \Psi({t}) \rangle = \prod_{\nu=1}^{N_e} \psi_\nu^\dagger({t}) \vert 0 \rangle$ where
$\psi_\nu^\dagger$ is represented by
$\psi_\nu^\dagger({t}) = \sum_{\mu=1}^{2N} \phi_\mu^\dagger({t}) u_{\mu\nu}({t})$,
and the unitary matrix $u_{\mu\nu}({t}) = \langle 0 \vert \phi_\mu({t}) \psi_\nu^\dagger({t}) \vert 0 \rangle$ satisfies the initial condition $u_{\mu\nu}({t}=0)=\delta_{\mu\nu}$.
During a short time interval $[{t},{t}+\delta{t}]$, $u_{\mu\nu}(t+\delta t)$ is obtained from $u(t)$ as
\begin{align}
u_{\mu\nu}({t}+\delta {t})
&= \sum_{\lambda=1}^{2N} \langle \mu({t}+\delta{t}) \vert \lambda({t})\rangle e^{i\varepsilon_\lambda({t})\delta{t}} u_{\lambda\nu}({t}) ,
\label{eq:sch}
\end{align}
where $\vert \lambda({t}) \rangle = \phi_\lambda^\dagger({t}) \vert 0 \rangle$.
The dynamics of the localized spins is described by the Landau-Lifshitz-Gilbert (LLG) equation,
\begin{align}
\frac{\partial\bm{S}_i}{\partial t} = \bm{h}_i^{\mathrm{eff}} \times \bm{S}_i + \alpha \bm{S}_i \times \frac{\partial\bm{S}_i}{\partial t} ,
\label{eq:llg}
\end{align}
where $\bm{h}_i^{\mathrm{eff}}({t}) = -\langle \partial \mathcal{H}({t}) /\partial \bm{S}_i \rangle = (J/S) \sum_{ss'} \langle c_{is}^\dagger \bm{\sigma}_{ss'}c_{is'}\rangle$ is the effective field, and $\alpha$ is the damping constant.
A small randomness is introduced in the localized spins at each site in the initial state;
the maximum deviation of the polar angle is $\delta\theta = 0.1$, which mimics the thermal fluctuation with temperature $T \sim 0.001$.
The (anti)periodic boundary condition is imposed on the electrons along the $x$ ($y$) direction.
The cluster size is chosen to $N=12\times 12$ and $16 \times 16$.
The time step is set to $\delta t = 0.005$ for $N=12\times 12$ and $\delta t = 0.01$ for $N=16\times 16$.
Other parameter values are $F_0=2$, $\mathit{\Omega}=1$, and $\alpha=1$.
The magnitude of the localized spins, $S$, is taken to be unity.
A unit of time, $t=1$, corresponds to $0.66$~fs for $h=1$~eV.

\begin{figure}[t]
\centering
\includegraphics[scale=1]{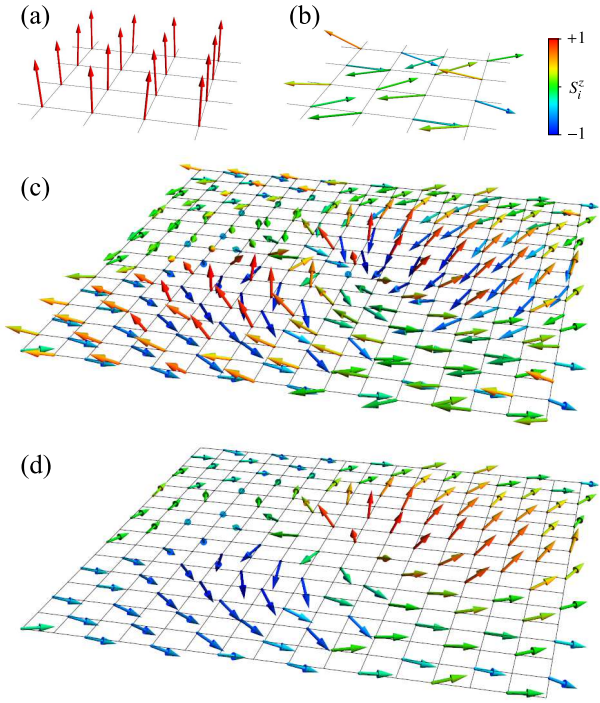}
\caption{(Color online)
Snapshots of the local-spin configurations at (a) $t=0$ (an initial FM state), (b) $t=1000$ (a steady AFM state) and (c) $t=500$.
A spin configuration in a sublattice at $t=500$ is shown in (d).
Color describes $S^z_i$.
The number of the lattice sites is $N=16\times 16$.
The periodic boundary condition is imposed on the localized spins.
}
\label{fig:snap}
\end{figure}

First, we show the transient spin structure in real space.
Figure~\ref{fig:snap} shows the snapshots of the local-spin configurations.
As reported in Ref.~\citen{ono1}, the initial FM order [Fig.~\ref{fig:snap}(a)] is collapsed by photoirradiation, and finally, almost perfect N\'eel state [Fig.~\ref{fig:snap}(b)] emerges.
We note that the spin anisotropy is not introduced explicitly in the Hamiltonian, and the directions of the magnetic orders are governed by the initial conditions.
A transient spin configuration during the FM to AFM orders is presented in Fig.~\ref{fig:snap}(c) where a non-coplaner and flow spin structures are confirmed.
A vortex-like topological structure is clearly shown in Fig.~\ref{fig:snap}(d), where spins are plotted in one of the two sublattices in a square lattice.

\begin{figure}[t]
\centering
\includegraphics[scale=1]{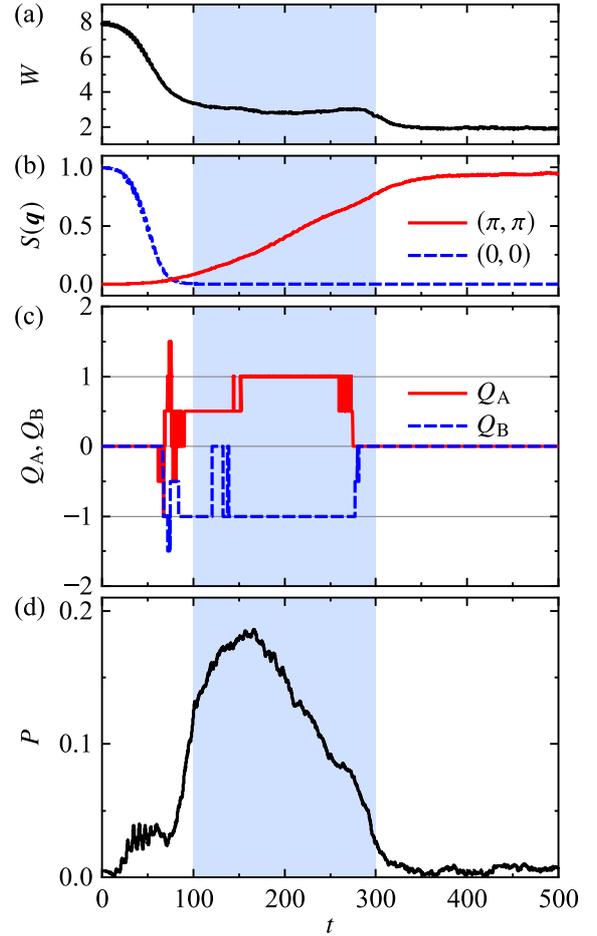}
\caption{(Color online)
Time profiles of (a) band width $W$, (b) FM spin structure factor $S(0, 0)$ and AFM one $S(\pi, \pi)$, respectively, (c) Pontryagin number $Q_{\rm A(B)}$ in each sublattice, and (d) vector chirality $P$.
Shaded areas indicate the ``intermediate time domain'' (see text).
The number of sites is $N=12\times 12$.
}
\label{fig:time}
\end{figure}

Real-time dynamics of the topological spin textures are analyzed in more detail.
We show the time dependences of the electron band width $W$ (Fig.~\ref{fig:time}(a)), and the FM spin structure factor $S(0, 0)$ and the AFM one $S(\pi, \pi)$ (Fig.~\ref{fig:time}(b)), which have been reported in Ref.~\citen{ono1}.
The spin structure factor is defined by $S(\bm{q}) = N^{-2} \sum_{ij} e^{i\bm{q}(\bm{r}_i-\bm{r}_j)} \bm{S}_i {\cdot} \bm{S}_j$.
There are two characteristic time scales: $t_{\rm F}\ (\sim 100)$ when the initial FM order almost collapses, and $t_{\rm AF}\ (\sim 300)$ when the steady AFM order is settled.
During $t_{\rm F}$ and $t_{\rm AF}$, shown by a shaded area in Fig.~\ref{fig:time}, a plateau emerges in the time profile of $W$, and $S(\pi, \pi)$ increases gradually.
This time region is termed an ``intermediate time domain'' from now on.

We calculate the Pontryagin index, i.e., the winding number, defined in each sublattice by
\begin{align}
Q_{\mathrm{A(B)}} 
&= \frac{1}{8\pi} \sum_{i\in \mathrm{A(B)}} \Bigl[ \mathcal{A}_{i,i+\hat{x}-\hat{y},i+\hat{x}+\hat{y}} + \mathcal{A}_{i,i+\hat{x}+\hat{y},i-\hat{x}+\hat{y}} \notag \\
&\quad + \mathcal{A}_{i,i-\hat{x}+\hat{y},i-\hat{x}-\hat{y}} + \mathcal{A}_{i,i-\hat{x}-\hat{y},i+\hat{x}-\hat{y}}
\Bigr],
\label{eq:qab}
\end{align}
where $\mathcal{A}_{ijk}$ is the solid angle subtended by three vectors, $\bm{S}_i$, $\bm{S}_j$, and $\bm{S}_k$.
The subscript $i+m\hat{x}+n\hat{y}$ represents the site at $\bm{r}_i+(m,n)$.
The summation in Eq.~\eqref{eq:qab} is taken for sites belonging to the sublattie A or B.
According to the spherical trigonometry, the solid angle $\mathcal{A}_{ijk}$ is evaluated as
\begin{align}
\mathcal{A}_{ijk}
= \sgn[\bm{S}_i\cdot(\bm{S}_j\times \bm{S}_k)] (C_{ijk}+C_{jki}+C_{kij}-\pi) ,
\end{align}
with
\begin{align}
C_{ijk} &= \cos^{-1} {\left [ \frac{\bm{S}_j\cdot\bm{S}_k-(\bm{S}_k\cdot\bm{S}_i)(\bm{S}_i\cdot\bm{S}_j)}{\sqrt{1-(\bm{S}_k\cdot\bm{S}_i)^2} \sqrt{1-(\bm{S}_i\cdot\bm{S}_j)^2}} \right]},
\end{align}
where $\sgn$ is the sign function.
We note that, when $\mathcal{A}_{ijk}$ is much smaller than $4\pi$, the solid angle is approximately given by the scalar product:
\begin{align}
\mathcal{A}_{ijk} \approx \chi_{ijk} \equiv \frac{1}{2} \bm{S}_i \cdot (\bm{S}_j \times \bm{S}_k).
\label{eq:scalarproduct}
\end{align}
It is shown in Fig.~\ref{fig:time}(c) that the sublattice winding numbers are finite during the intermediate time domain, i.e., $Q_{\mathrm{A}}=-Q_{\mathrm{B}}=1$, indicating the appearance of the topological defects.
We also examine the uniform vector chirality defined by
\begin{align}
P=\frac{1}{N} {\left\vert \sum_{\langle ij \rangle} \bm{e}_{ij} \times
(\bm{S}_i \times \bm{S}_j) \right\vert},
\end{align}
where $\bm{e}_{ij} \propto (\bm{r}_j-\bm{r}_i)$ is the unit vector directing from site $i$ to site $j$.
This quantity is known to reflect the electric polarization induced by the chiral spin structure.~\cite{katsura}
As shown in Fig.~\ref{fig:time}(d), $P$ is remarkable in the intermediate time domain, although it does not vanish even in $t<t_{\rm F}$ and $t>t_{\rm AF}$.

\begin{figure}[t]
\centering
\includegraphics[width=1\columnwidth,clip]{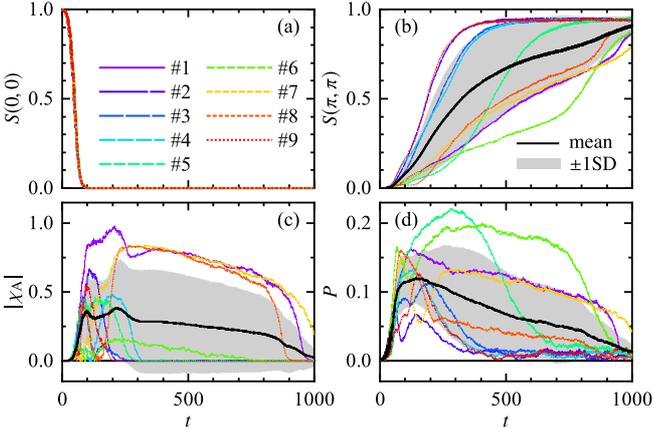}
\caption{(Color online)
Time profiles of (a)~spin structure factor at $(0, 0)$, (b)~that at $(\pi, \pi)$,
(c) scalar chirality in the sublattice A, and (d) vector chirality, for different initial spin configurations (\#1--\#9).
Bold curves and shaded areas represent the mean values and the standard deviations, respectively.
The number of the lattice sites is $N=16\times 16$.
}
\label{fig:initial}
\end{figure}

Characteristics of the topological spin structure depend on the initial spin structure.
As explained previously, a small randomness is introduced in $\bm{S}_i$'s at each site in the initial state.
Numerical simulations for the real-time evolution are carried out for different 9 initial states, and the mean values and standard deviations of the time profiles of physical quantities are presented in Fig.~\ref{fig:initial}.
Here, we introduce the staggered scalar chirality instead of the Pontryagin index as
\begin{align}
\chi_{\mathrm{A(B)}} &= \frac{1}{8\pi} \sum_{i\in \mathrm{A(B)}} \Bigl[ \chi_{i,i+\hat{x}-\hat{y},i+\hat{x}+\hat{y}} + \chi_{i,i+\hat{x}+\hat{y},i-\hat{x}+\hat{y}} \notag \\
&\quad + \chi_{i,i-\hat{x}+\hat{y},i-\hat{x}-\hat{y}} + \chi_{i,i-\hat{x}-\hat{y},i+\hat{x}-\hat{y}} \Bigr],
\end{align}
where the summation is taken over the sublattice A(B).
Data sets of $S(0, 0)$ shown in Fig.~\ref{fig:initial}(a) obtained from the different initial states are located on a universal curve.
On the other hand, other three quantities depend largely on the randomness introduced in the initial states.
Bold black lines and shaded areas in Fig.~\ref{fig:initial} represent the mean values and the standard
deviations, respectively.
The time profiles are classified qualitatively into the two types.
Type~I: $S(\pi, \pi)$ increases rapidly and saturates smoothly to one.
Type~II: a plateau-like features appear in the time profiles of $\chi_{\rm A}$ and $P$, although the plateau values and their time regions show some varieties.
The time profiles of $\chi_{\rm A}$ and $P$ show peak structures when $S(\pi, \pi)$ increases steeply in type~I, and show plateau structures in type~II.
The results imply that the photoinduced topological objects are metastable, and the N\'eel state is settled after the topological objects fade.

\begin{figure}[t]
\centering
\includegraphics[scale=1]{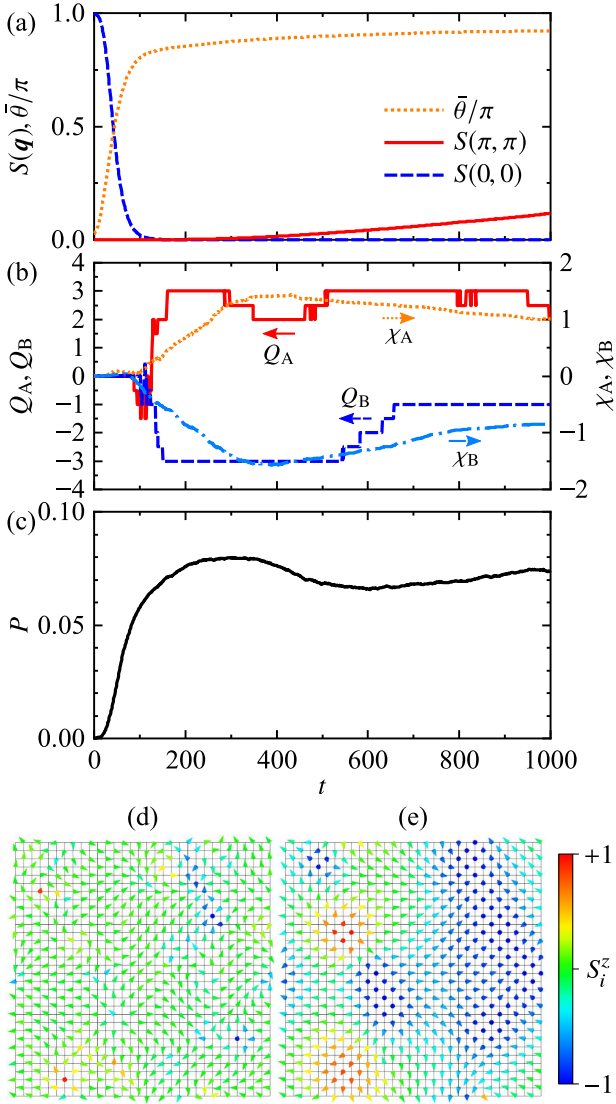}
\caption{(Color online)
Time profiles of (a) spin structure factors $S(\bm{q})$ and nearest-neighbor spin angle $\bar{\theta}$, (b) Pontryagin index $Q_{\rm A(B)}$ (bold and broken lines), and scalar chirality $\chi_{\rm A(B)}$ (dotted and dashed-dotted lines), and (c) vector chirality $P$ calculated by the kernel polynomial method.
(d)(e) Snapshots of the localized spin configurations in the sublattice A at (d)~$t=200$ and (e)~$t=1000$.
Color represents $S^z_i$.
The chemical potential of the electrons is changed from $\mu_i = -J/S = -4$ to $\mu_f = 0$ at $t=0$ corresponding to the sudden quench from the equilibrium FM metallic to AFM insulating states.
The number of the lattice sites is $N=32\times 32$.}
\label{fig:kpm}
\end{figure}

The emergence of the topological objects shown above is checked by a different method where numerical calculations in large size clusters of the order of $10^4$ are possible.
We analyze the quench dynamics from the FM to AFM orders by using the kernel polynomial method combined with the automatic differentiation technique,~\cite{weibe,barros,ozawa,wang} where the conduction electrons are assumed to be in equilibrium with temperature $T$.
When the localized spins are treated as classical vectors, the Hamiltonian in Eq.~\eqref{eq:hamiltonian} and the partition function are written as $\mathcal{H} = \sum_{ijss'} c_{is}^\dagger h_{is,js'}[\phi] c_{js'}$ and $Z=\Tr_{\phi} \Tr_{c} \exp[-\mathcal{H}/T]$, respectively, where $\phi=\{\bm{S}_i\}$ is the classical spin configuration.
Taking the trace over the electron degree of freedom denoted by $\Tr_c$, we obtain the free energy as $\mathcal{F}[\phi] = -T \int d\omega\, \rho(\omega) \ln[1+e^{-(\omega-\mu)/T}]$, where $\rho(\omega)=\sum_\nu \delta(\omega-\varepsilon_\nu)$ is the single-particle density of states with the $\nu$th eigenvalue $\varepsilon_\nu$ of $h_{is,js'}[\phi]$.
The kernel polynomial method and the automatic differentiation technique enable one to evaluate efficiently the free energy $\mathcal{F}[\phi]$ and the effective field $\bm{h}_i^{\mathrm{eff}}=-\partial \mathcal{F}[\phi]/\partial \bm{S}_i$ with a computational time that scales as $\mathcal{O}(N)$.
We consider the sudden quench of the chemical potential from $\mu_i=-J/S$ to $\mu_f=0$, which correspond to the equilibrium FM metallic ($n_e=0.5$) and AFM insulating $(n_e=1)$ states, respectively.
The number of the random vectors for the stochastic estimates of the trace is $100$, and the Chebyshev expansion is taken up to the 200th order.
The number of the sites and the time step are $N=32\times 32$ and $\delta t = 0.1$, respectively, and the other parameter values are the same as before.

Figure~\ref{fig:kpm}(a) shows time profiles of the spin structure factors $S(\bm{q})$, and the nearest-neighbor spin angle defined by $\bar{\theta} = (2N)^{-1} \sum_{\langle ij \rangle} \cos^{-1}(\bm{S}_i\cdot\bm{S}_j)$.
The Pontryagin index and the staggered scalar chirality are plotted in Fig.~\ref{fig:kpm}(b), and the uniform vector chirality $P$ is presented in Fig.~\ref{fig:kpm}(c).
Real space spin configurations at $t=200$ and $1000$ are presented in Figs.~\ref{fig:kpm}(d) and (e), respectively.
After the quench of the chemical potential, the FM correlation $S(0,0)$ decreases and $\bar{\theta}$ increases rapidly until $t \sim 100$.
The uniform vector chirality develops similarly to $\bar{\theta}$.
Then, in $t \gtrsim 100$, $\bar{\theta}$ and $P$ remain almost constant within the present upper limit of the timescale, which is supposed to be smaller than $t_{\rm AF}$.
Figure~\ref{fig:kpm}(d) shows a snapshot of the localized spins at $t=200$, where seeds of the topological defects represented by red and blue (black in grayscale) cones are seen.
In the intermediate time domain of $t\gtrsim 100$, the sublattice Pontryagin index and the staggered scalar chirality appear as in Figs.~\ref{fig:time} and \ref{fig:initial}.
Similarly to Fig.~\ref{fig:time}(c), the Pontryagin indices $Q_{\rm A}$ and $Q_{\rm B}$ take not only integer values but also half-integer values, and do not always satisfy $Q_{\rm A}=-Q_{\rm B}$.
Thus, the topological defects are interpreted as meron- or vortex-like objects.
The slope of the time profile of $S(\pi,\pi)$ is much smaller than that in the $N=12\times 12$ lattice in Fig.~\ref{fig:time}(b) and that in the $N=16\times 16$ lattice in Fig.~\ref{fig:initial}(b), implying metastability of the topological objects in the thermodynamic limit.
We conclude that the emergence of the topological spin objects is attributed to the sudden change in the initial FM order, and is irrespective of the detail of the conduction-electron dynamics.

The present numerical calculations reveal that the topological spin textures are produced through the photoinduced spin-structure change from the FM to AFM orders.
We propose the following interpretation of the observations.
The initial FM state is first collapsed by light irradiation around $t=t_{\rm F}$, and the system is in a moment a paramagnetic state.
Then, the exchange interaction is transformed into the AFM one in highly non-equilibrium state, in contrast to the conventional FM DE interaction, and spins tend to align antiferromagnetically.
In the transient state ($t_{\rm F}<t<t_{\rm AF}$) before the perfect N\'eel state is realized, the AFM domains with different directions of the staggered magnetization grow inside of the paramagnetic region.
Around the domain boundaries, mismatches of the spin alignments happen, which produces seeds of the topological spin defects.
This is a realization of the Kibble-Zurek mechanism for the topological defect generation in the phase transition dynamics.~\cite{kibble,zurek,dziarmaga}

In conclusion, the photoinduced topological spin textures are examined in the DE model.
In the transient non-equilibrium state from the initial FM to steady AFM states,
the Pontryagin index, the scalar and vector chiralities emerge, indicating the photoinduced topological spin textures.
This production is attributed to the dynamical change of the magnetic long-range ordered states.
The present study provides a new route to produce the topological spin texture in magnets by using the laser light without SOI.

\begin{acknowledgment}
We thank T.~Oka, G.~Tatara, S.~Iwai, and S.~Koshihara for their helpful discussions.
This work was supported by JSPS KAKENHI Grant No.~JP15H02100, JP17H02916, JP18H05208 and JP18J10246.
Some of the numerical calculations were performed using the facilities of the Supercomputer Center, the Institute for Solid State Physics, the University of Tokyo.
\end{acknowledgment}

\end{document}